\documentclass{elsart}

\usepackage{amssymb}
\usepackage{latexsym}
\usepackage{amsmath}
\usepackage{amsfonts}

\usepackage{epsfig}

\begin{document}
\journal{Physica A}
\begin{frontmatter}

\title{A family-network model for wealth distribution in societies}

\author[1]{Ricardo Coelho},
\author[1,2]{Zolt\'an N\'eda\corauthref{cor1}}\corauth[cor1]{} 
\ead{zneda@phys.ubbcluj.ro}, 
\author[1,3]{Jos\'e J. Ramasco}
\and \author[1]{Maria Augusta Santos}

\address[1]{Departamento de F\'{\i}sica and Centro de F\'{\i}sica do Porto,
Faculdade de Ci\^encias, Universidade do Porto, Rua de Campo Alegre 687,
4169-007 Porto, Portugal}

\address[2]{Dept. of Physics, Babe\c{s}-Bolyai University,
          str. Kog\u{a}lniceanu 1, RO-400084 Cluj-Napoca, Romania}

\address[3]{Physics Department, Emory University, Atlanta, Georgia 30322, USA}

\date{November 2004}

\begin{abstract}

A model based on first-degree family relations network is used to describe the
wealth distribution in societies. The network structure is not a-priori introduced in 
the model, it is generated in parallel with the wealth values through simple and realistic 
dynamical rules. The model has two main parameters, governing the wealth exchange in the
network. Choosing their values realistically, leads to wealth distributions in good
agreement with measured data. The cumulative wealth distribution 
function has an exponential behavior in the low and medium wealth limit, 
and shows the Pareto-like power-law tail for the upper $5\%$ of the society. The
obtained Pareto indexes are in good agreement with the measured ones. The generated 
family networks also converges to a statistically stable topology with a simple Poissonian degree  distribution. On this family-network many interesting correlations are studied, and the
main factors leading to wealth-diversification and the formation of the Pareto law are identified. 

\end{abstract}


\end{frontmatter}

\section{Introduction}

Since the seminal work by Vilfredo Pareto\cite{pareto}, it is
known that the wealth distribution in capitalist economies shows a
very peculiar and somehow universal functional form. In the range 
of low income, the cumulative distribution of wealth (the probability that the
wealth of an individual is greater than a given value) may be fitted by an 
exponential or log-normal decreasing function, while in the region containing 
the richest part of the population, generally less than the 
$5\%$ of the individuals, this distribution is well characterized 
by a power-law (see for example \cite{ebooks} for a review). This empirical
behavior has been confirmed by a number of recent studies on the
economy of several corners of the world. The available 
data is coming from so far apart as Australia \cite{matteo03}, Japan
\cite{aoyama03,fujiwara03}, the US \cite{silva04}, continental
Europe \cite{fujiwara04,clementi04} or the UK \cite{dragulescu01}. The data
is also spanning so long in time as ancient Egypt \cite{abul02},
Renaissance Europe \cite{souma02} or the $20th$ century
Japan\cite{souma01}. Most of these data are based on the declaration of income of the population,
which is assumed to be proportional to the wealth. There are however some 
other databases obtained from different sources like for instance the area 
of the houses in ancient Egypt \cite{abul02}, the inheritance taxation 
or the capital transfer taxes \cite{UK}. The results mostly back 
Pareto's conjecture on the shape of the wealth distribution. 
The interesting problem that remains to be answered is the origin of the
peculiar functional trend.

The answer to this question is a long standing problem, which even
motivated some of the initial Mandelbrot's and Simon's work fifty
years ago. Let $P_>(w)$ be the probability of having a wealth
higher than $w$. Pareto's law then establishes that the tail
of $P_>(w)$ decays as
\[
P_>(w) = \int_w^\infty P(w') \, dw' \sim w^{-\alpha} ,
\]
where $\alpha$ is the so called Pareto index and $P(w)$ the normalized wealth
distribution function. Typically, the
presence of power-law distributions is a hint for the complexity
underlying a system. It is however important to notice that in
spite of what happens with most exponents in Statistical Physics
$\alpha$ may change in time depending on the economical
circumstances \cite{fujiwara03,souma01}, making thus impossible
the definition of some sort of universal scaling in this problem. 
This aspect is a key characteristic that any model on wealth
distribution should be able to reproduce.

Economical models are essentially composed by a group of agents
placed on a lattice that interchange money following
pre-established rules. The system will eventually reach a
stationary state where some quantities, as for
instance the distribution $P_>(w)$, may be measured. Following
these ideas, Bouchaud and M\'ezard \cite{bouchaud00} and Solomon and
Richmond \cite{biham98,solomon01} separately proposed a very
general model for wealth distribution. This model is based on a
mean field type scenario with interactions among all the agents
and on the existence of multiplicative fluctuations acting on each
agent's wealth. Their results on the wealth distributions 
adjust well to the phenomenological $P_>(w)$.
Roughly the same conclusions were obtained by Scaffeta \cite{scafetta04},
who considered a nonlinear version of the model and from other regular 
lattice based models as those in Refs. \cite{chatterjee03} and \cite{pianegonda03}. 
This kind of models defined on pre-established regular lattices is 
however unable, by construction, to account for the complexity of 
the interaction network observed in real economical systems.

In parallel to the previous efforts to characterize economical 
systems, the study of complex networks has experienced a burst of 
activity in the last few years (see Ref. \cite{dorogovtsev03} 
for a recent review). Social networks,
in particular, are of paramount interest for economy since
everyday economical transactions actually produce 
a network of this type. The topology of the economical network can indeed
condition the output of any economical model running on it. Such
effect has been documented for example in Refs.
\cite{souma01,iglesias03,garlaschelli04}, in which the models
described above were simulated on Small-World or Scale Free
networks. One of the main characteristics of social networks is
the positive correlation existing between the node degrees
\cite{newman03,ramasco04}, i.e., the high connected individuals
commonly tend to connect with other well connected people. The way
of constructing this type of networks is precisely the main topic
of a recent work by Bogu\~na and coworkers \cite{bogunya04}. In
what follows, we are going to use a somehow similar approach to grow our
working network.

Bogu\~na's method is based on the existence of hidden variables
characterizing each agent state. In this work, and in the spirit
of Ref. \cite{garlaschelli04b}, we present a simple economic model where
those hidden variables are identified with the wealth of each
agent. This introduces a coupling between the dynamics of the network
structure and the evolution of the wealth distribution. Each value 
of the external parameters thus determines not only the final wealth 
distribution but also the structure of the underlying interchange 
network. 

This paper is organized as follows. In section 2 we introduce our model,
in which the agents are identified as families linked by
first-degree family relationship. In section 3 we present
computational results on this model. For a wide range of the 
parameters of the model we study both the wealth distribution and the structure of the
underlying network. In section 4 we discuss our results from several viewpoints.
In this section the results are compared with real data on wealth 
distribution, the correlation between the wealth and connectivity of the agents is studied, 
and the dynamics leading to wealth diversification is investigated. 
Section 5 is then dedicated to conclusions.

\section{The family-network model}

In modeling the wealth distribution in societies we identify as main entities 
(agents) the families. In the framework of our model, the families are nodes 
in a complex network, and the links of this network are first-degree family relations.
Beside its links, each node is characterized by its "age", $A(i)$, and wealth, $w(i)$. 
The age of a node is proportional to the simulation time-steps
elapsed from its birth ($A(i)=t-t_b(i)$, where $t_b(i)$ denotes the time-step when node $i$
was born), and the wealth is a positive real number that will change in time. 
We consider both the total wealth of the system, $W_t$, and the number of families (nodes),
$N$, conserved. The structure of the network is not a-priori fixed, and 
will also change during the evolution of the system. Initially we start with nodes arranged on a 
regular hierarchical network (as sketched in Figure 1) where the age of node $i$ is
simply $N+1-i$. In this manner node $1$ will be the oldest and node $N$ the youngest one.
It is worth mentioning here that the final statistics of wealth distribution and
the final network topology are rather independent on how the initial network
topology was chosen. We verified this by choosing several other qualitatively different 
initial network structures.

\begin{figure}
\begin{center}
\epsfysize=50mm
\epsffile{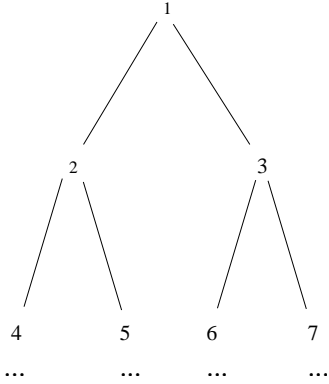}
\caption{Initial structure of the network}
\label{fig1}
\end{center}
\end{figure}

Initially we also assign random wealth to each node according to a uniform distribution
on the $(0,1)$ interval. In this manner we constructed the start-up society with a simple 
network structure (family relationships) and randomly distributed wealth values. 
The time-evolution of the system is then chosen to be as simple as possible, but capturing the
realistic wealth exchange processes between families. For each simulation time-step the
dynamics is as follows:
\begin{enumerate} 
\item The oldest node (let this be $j$) is taken away from the system. 
The wealth of this node is uniformly redistributed between 
its first neighbors (nodes that are linked to it), and all its links are deleted.  
\item Node $j$ is re-introduced in the network with age $A(j)=0$. 
It is linked to two randomly selected nodes (let these be $k$ and $l$) that have 
wealth greater than a minimal value $q$. The wealth $q$ is taken away from the wealth of the selected $k$ and $l$ nodes, and it is redistributed in a random and preferential manner in the society. The preferential redistribution is realized by splitting the $2q$ wealth in $s$ parts and choosing the the nodes which will benefit from these parts with a probability proportional to their actual wealth. This preferential redistribution  will favor a rich-get-richer effect. After the redistribution of the $2q$ amount, a $p$ part ($p<1$) of the remaining wealth of node $k$ and $l$ is given as start-up wealth for node $j$. After these wealth redistribution
processes the wealth of node $k$ and $l$ will be thus $w'(k)=[w(k)-q](1-p)$ and $w'[l]=
[w(l)-q](1-p)$, respectively. Node $j$ will start with $w(j)=p[w(k)+w(l)-2q]$ wealth.  
\item The age of all nodes is increased by unity.
\end{enumerate}

Let us now explain the socio-economic phenomena that are modeled by the above dynamics. 
Step 1 models the inheriting process following the death of one family. The wealth of
this family is redistributed among its first-degree relatives (children). Step 2 models the
formation of a new family. In order to create a new family two other families have to raise one
child. For raising a child a minimum amount of wealth is needed ($q$). This cost is paid to the society (for food, clothes, services...), and the members of the society will benefit unevenly from it. Families with bigger wealth control more business, so they will naturally benefit more. 
The preferential redistribution of the $2q$ wealth models this uneven profit, and it is
the main ingredient necessary to reproduce the Pareto distribution. Finally, when a new
family is born it is linked by first-degree relations to two existing families and 
gets a given part ($p$) of the parents wealth as start-up money. 
The time-scale of the simulation is governed by the time needed to change all nodes, which
we call one generation or one Monte Carlo step (MCS).  
By fixing $N$ and $W_t$, and studying the thermodynamic limit
$N \rightarrow \infty$, the model becomes essentially a two-parameter model ($q$ and $p$), 
which is suitable for extended computer simulations. 
 
Although very simple in nature, the chosen dynamics incorporates, we believe, 
the main socio-economic
factors that influence the redistribution of wealth between families. As time passes the families will be able to gather more and more wealth due to the $2q$ wealth redistribution process in the society. When their wealth becomes big enough they can create new families, and donate a part of this wealth to the new family.
This process is costly and will therefore lower their wealth. Very poor nodes will not likely reach the $q$ threshold and will not be able to create new families, becoming isolated nodes. There is no clear
determinism however, since the redistribution in step 2 is realized in a random manner, and the
selection of the two nodes to which the new family links is also random. So in
principle there is the chance that nodes that start with low wealth will
become very rich, or rich nodes do not increase their wealth as expected. The actual way
how the preferential redistribution in step 2 is implemented is by dividing the $2q$ value 
in many (usual several hundred) equal parts, and each part is assigned to a randomly chosen node, 
biased proportionally with the wealth of the node. To do this biased redistribution the use of 
a BKL \cite{BKL} type Monte Carlo algorithm is very helpful. Another possibility 
(leading to the same results) for doing this preferential redistribution would be
to select $s$ nodes with the same probability, independently
of their wealth, and then to split the $2q$ amount between 
the selected nodes proportionally with their actual wealth. 
It is also important to note that in realizing step 1, one can get to a situation where the selected node has no links (a family dies out 
without children). For simplicity reasons, 
in this case we have also chosen to redistribute the wealth of the node 
in the whole society by using the same preferential rule.  

Of course, this model is a rough description of the reality and it should be viewed only as a first "mean-field" approximation. In real societies the number of families and also 
the total amount of wealth should not be considered fixed.  
Many other social aspects could be of interest, the actual value of $q$
and $p$ should vary from family to family within quite broad distributions, 
the nodes must not die out according to their
age and many cultural and religious factors can influence the dynamics of the underlying
social network. In spite of all 
the neglected effects we will see that this simple model is able to
reproduce the observed wealth-distributions and generates reasonable first-degree
family relation networks. The main advantage of this model is that the network structure on which the
wealth-exchange is realized is not a-priori put in the system. The network forms and 
converges to a stable topology in time, together with the wealth diversification in the system and
the appearance of the Pareto distribution.       

\section{Results of the model}

Extensive computer simulations were done to study the wealth distribution and the generated 
family-network for various values of the model parameters. In order to minimize
the statistical fluctuations  we averaged over $100$ realizations for each parameters values. The model as defined above has several 
parameters: $N$, the number of nodes in the network, $W_t$, the total wealth of the system, $t$ the number of simulation steps done to reach a given state, the number $s$ giving the parts
on which the $2q$ wealth is divided, and the value of the $q$ and $p$ wealth-exchange parameters. By simple simulations it is easy to show that the results are independent of
the chosen value of $s$, provided that $s$ is big enough ($s\ge 10$ gives already stable
results). In the results that will be presented we always used $s=100$. We will argue in the following that the main free parameters are $q$ and $p$, since the model converges rapidly both as a function of time and as a function of the number of nodes to a stable limiting distribution and network-structure. 

It is easy to realize that the chosen value of $W_t$ will not change the nature of the results, but
it simply rescales the values of the wealth. A simple computer exercise will also convince us 
that the above defined family-network model converges in time very quickly to a statistically 
invariant state both for the wealth distribution and network structure. Results
for a relatively big lattice ($N=10000$) and for realistic $p=0.3$ and $q=0.7$ values
are presented in Figure 2. We see that roughly after 5 MCS, both the cumulative wealth distribution and the first two moments of the
degree distribution converge to their stable limit. 

\begin{figure}
\begin{center}
\epsfysize=70mm
\epsfxsize=130mm
\epsffile{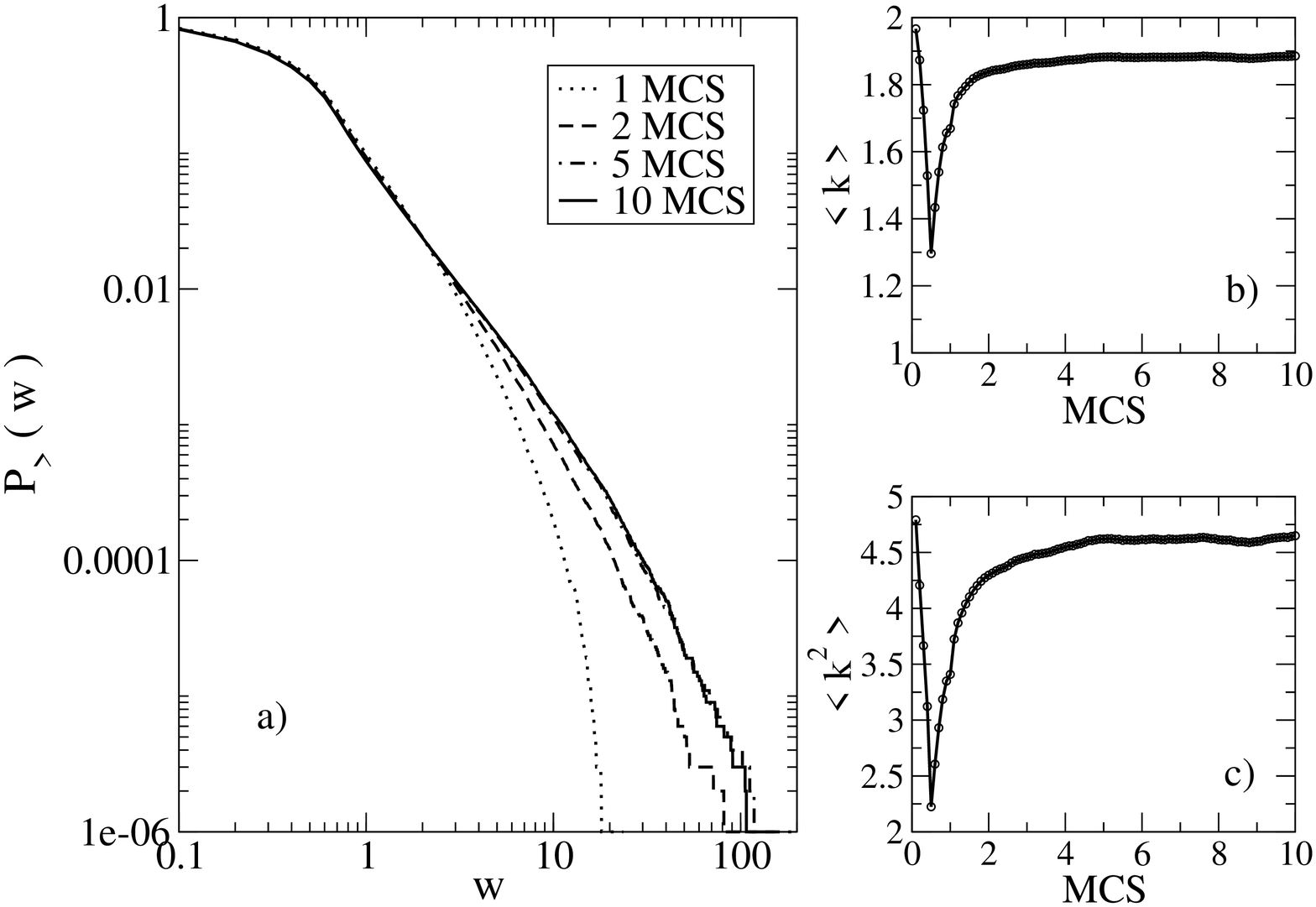}
\caption{Time evolution of the cumulative wealth distribution function (a.), 
average degree of the nodes (b.), and average square of the degree of the nodes (c.).
Simulation were done on a network with $10000$ nodes, $p=0.3$ and $q=0.7$.}
\label{fig2}
\end{center}
\end{figure}

On the other hand one can also check that the 
model has a well defined thermodynamic limit. As $N$ increases, we obtain again that
both the cumulative wealth distribution and the statistical properties of the network reach
a stable limit. Characteristic results for this variation are presented in Figure 3. As
we can see from the figure, for reasonably big lattices $N\approx 10000$, a stable limit is reached.

\begin{figure}
\begin{center}
\epsfysize=70mm
\epsfxsize=130mm
\epsffile{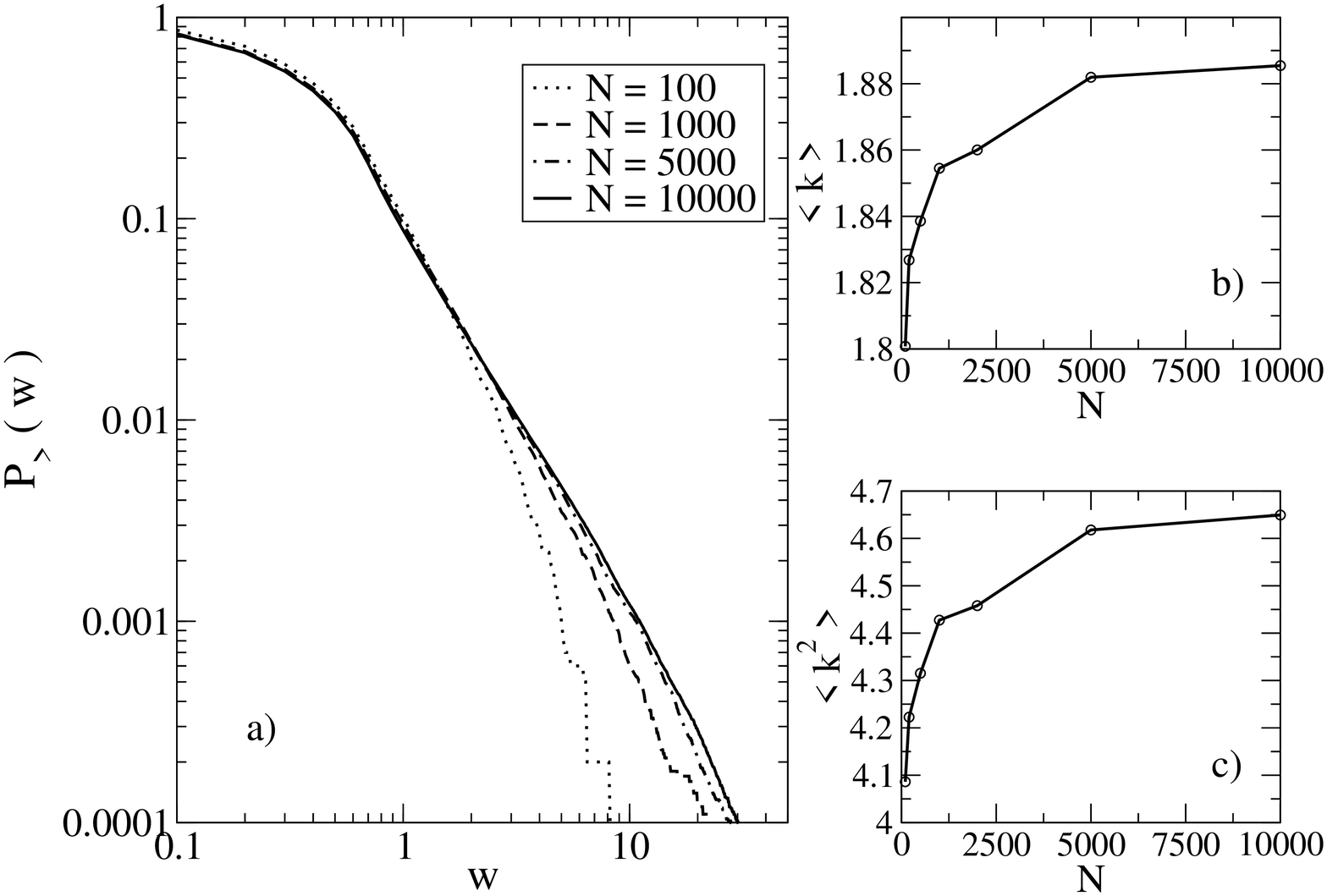}
\caption{Effects of the network size on the final results. The  stable (after $10$ MCS) cumulative wealth distribution function (a.), average degree- (b.), and
average square degree of the nodes (c.), all for different network sizes. Simulations
with $p=0.3$ and $q=0.7$.}
\label{fig3}
\end{center}
\end{figure}

We will study now the influence of the $p$ and $q$ wealth-exchange parameters. Since we have
verified that the model converges relatively quickly to a stable limit, we will
consider in all simulations 10 MCS. The number of nodes in the network will be chosen $N=10000$,
which ensures that the thermodynamic limit is approached. The $p$ parameter can theoretically vary in the $(0,1)$ interval; we
consider however, realistic a variation in the $(0.1-0.3)$ interval. 
Since start from wealth values distributed randomly and uniformly on the $(0,1)$ interval, the minimal
$q$ value needed to raise a child should thus also be in the $(0,1)$ interval, otherwise no
new family could be linked to the network.  First we present our results on the $P_>(w)$ cumulative wealth distribution curves. For two fixed values of $p$ ($p=0.1$ and $p=0.3$) the curves are given in Figure 4.

\begin{figure}
\begin{center}
\epsfysize=50mm
\epsfxsize=130mm
\epsffile{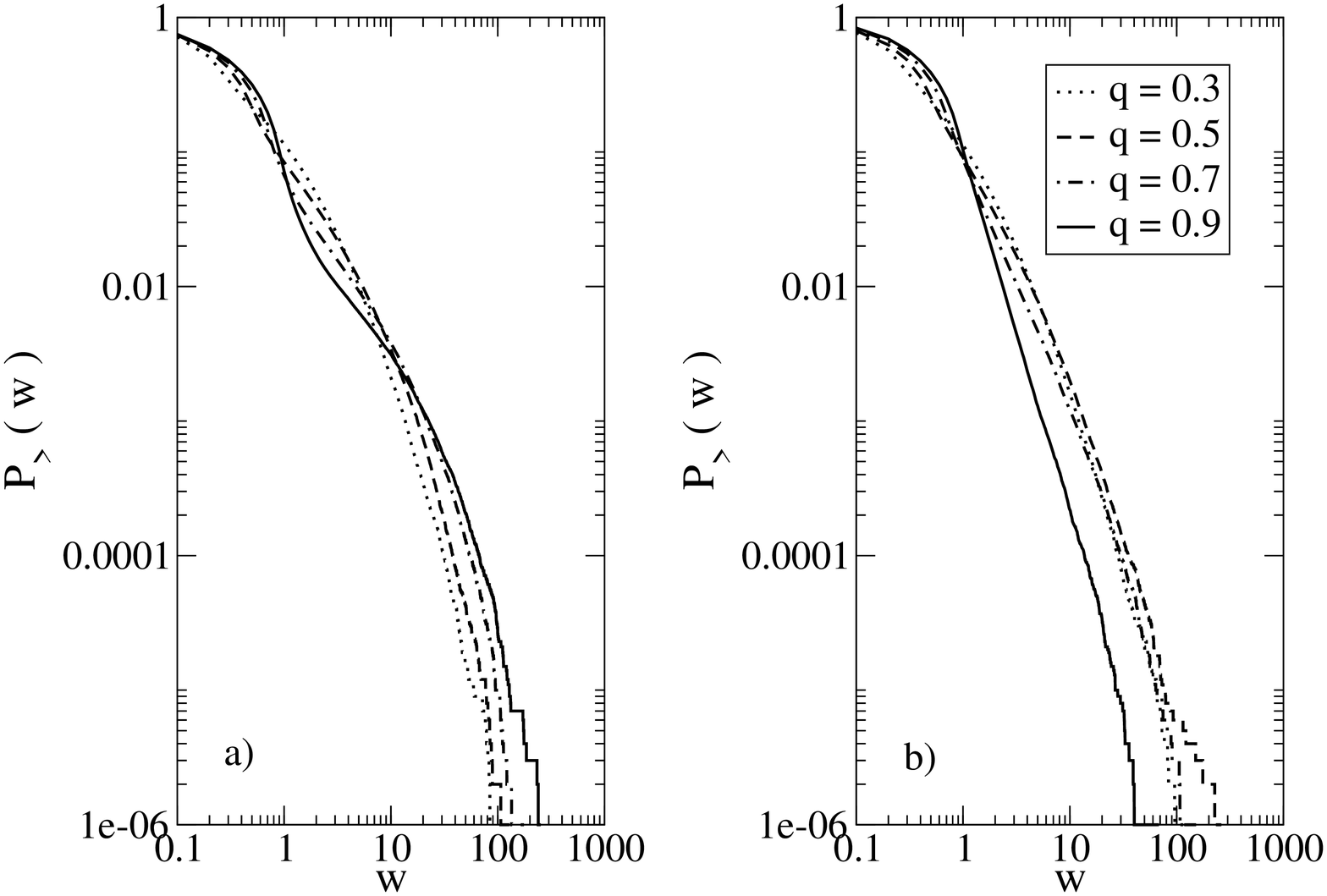}
\caption{Cumulative distribution functions for different $p$ and $q$ values, 
(a) $p=0.1$ and (b) $p=0.3$. Different curves are
for different $q$ values, as sketched on the legend of (b). The results are after
$10$ MCS and for $N=10000$.}
\label{fig4}
\end{center}
\end{figure}

The curves in Figure 4 suggest that the good scale-free Pareto tail is obtained for $q$ values in
the $(0.7-0.9)$ interval, and we will thus focus in the following on this parameter region. 
It is also evident that results for $p=0.3$ have a better trend. The Pareto
index (power-law exponent) in this region varies in the $(1.7-2.5)$ interval, depending 
on the chosen $p$ and $q$ values and fitting intervals. The $P_>(w)$ curves have the right shape, they show the power-law trend for the rich nodes and the exponential behavior in the low and medium wealth limit (Figure 5). Moreover, one can also observe that in good agreement with the reality, roughly $5-10\%$ of the nodes have wealth in the Pareto regime.

\begin{figure}
\begin{center}
\epsfysize=70mm
\epsffile{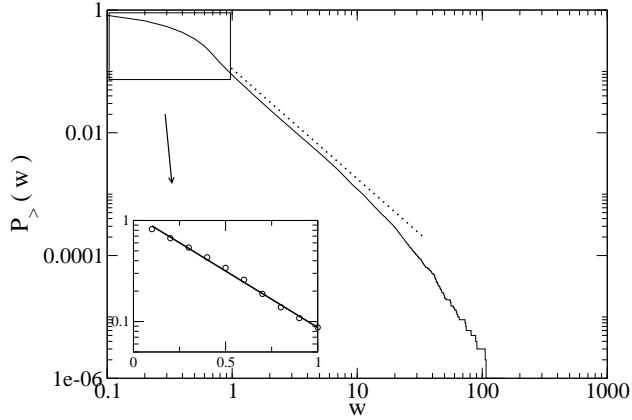}
\caption{The shape of the obtained cumulative wealth distribution function for 
$p=0.3$, $q=0.7$ ($N=10000$ and results after $10$ MCS). The tail is approximated
by power-law with exponent $\alpha=1.80$, and the initial part of the curve
has an exponential trend. The inset shows this initial trend on log-normal scale}
\label{fig5}
\end{center}
\end{figure}

The network generated by the model is a simple exponential one. 
Considering the realistic $q \in (0.7-0.9)$ and $p \in (0.1-0.3)$ parameter region, in Figure 6
we present results obtained for the $P(k)$ degree distribution (probability-distribution that one
node has a given number of links). From the degree distribution we conclude that the network is an exponential one. The most probable connectivity of a node is around $2$, and we obtained that in this parameter region $\langle k \rangle$ varies between $1.8-1.9$, which are reasonable values for real first-degree family relation networks. No relevant clusterization was observed in these networks. 

\begin{figure}
\begin{center}
\epsfysize=70mm
\epsffile{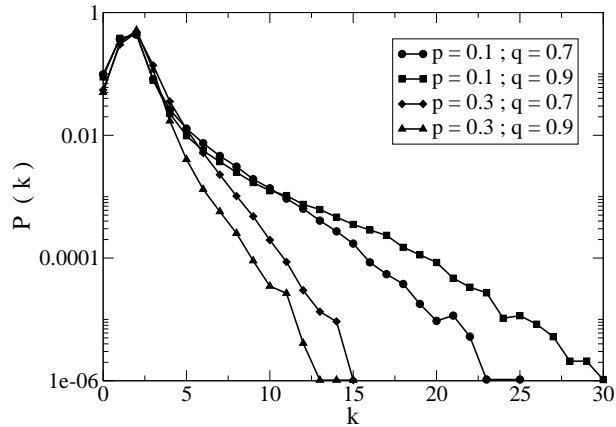}
\caption{The degree distribution of the obtained networks on a log-normal scale
for various values of $q$ and $p$. (Simulations after $10$ MCS and with $N=10000$ 
nodes)}
\label{fig6}
\end{center}
\end{figure}

It is also instructive to study different kinds of correlations in the generated networks. First
one can study the correlation between the $k$ connectivity of the nodes, and mean-connectivity of the neighbors $\langle k_{nn} \rangle$ for nodes with $k$ links. If there exist a positive
degree-degree correlation, i.e. if well connected nodes tend to connect with well connected ones,
then $\langle k_{nn}(k) \rangle$ must increase with $k$.  
In the relevant parameter region results in this sense, are plotted in Figure 7. For low values of $p$ there are no 
obvious correlations, but as 
$p$ increases one can observe a positive correlation effect,
$\langle k_{nn}(k) \rangle$ increases roughly linearly with $k$. 
This means that, if the new family 
gets a bigger portion of the parents wealth, the number of links parents and children have are
positively correlated. The effect is simply understandable, taking into account that 
for higher values of $p$ the wealth of the parents and children should be also correlated, 
creating similar conditions for accepting links.

\begin{figure}
\begin{center}
\epsfysize=70mm
\epsffile{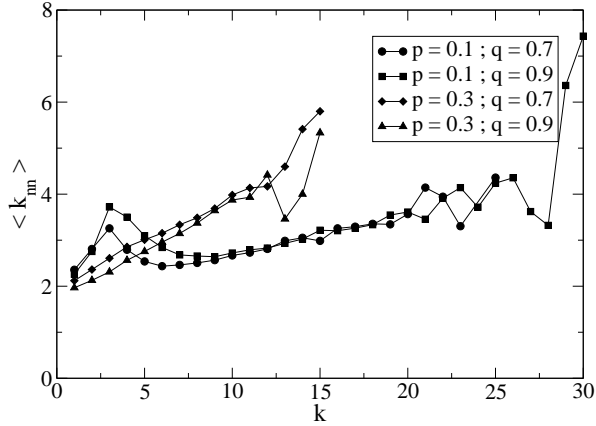}
\caption{ Mean connectivity of the neighbors ($ \langle k_{nn} \rangle$) as a function of the
connectivity of the nodes for various values of $q$ and $p$. (Results
obtained after $10$ MCS and $N=10000$.)}
\label{fig7}
\end{center}
\end{figure}

The correlation between the wealth $w$ of one node and the 
average wealth of the neighbors $\langle w_{nn} \rangle$, 
should follow a similar trend. Indeed, as 
expected, this correlation also has an increasing trend as
$p$ is increased (Figure 8a). This positive correlation effect is more 
clear again for not too high wealth values, since in the high $w$ limit there are few 
nodes and the statistics is poor. A similar correlation trend can be observed if 
one studies the correlation between the wealth of the nodes and the 
total wealth of the neighbors. In Figure 8 
we plotted the results only for $w \le 5$, since for higher values of 
$w$ the curves are very noisy due to the poor statistics.

\begin{figure}
\begin{center}
\epsfysize=70mm
\epsfxsize=100mm
\epsffile{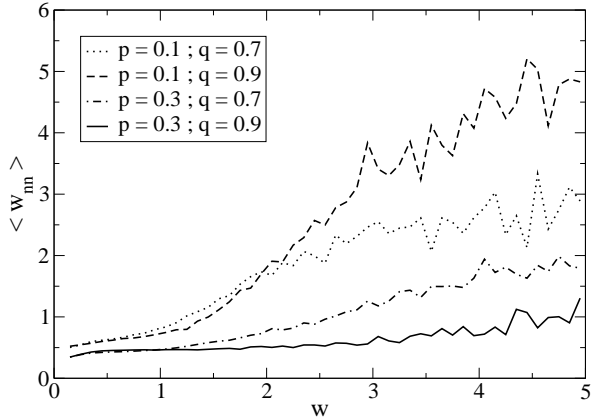}
\caption{ Average wealth of the neighbors as a function of the wealth of the
nodes. Different values of the $q$ and $p$ parameters are considered. 
(Results after $10$ MCS and for $N=10000$.) }
\label{fig8}
\end{center}
\end{figure}

Finally, one can study the correlation between the wealth and connectivity of a node, either
by plotting $\langle k (w) \rangle$ (the average number of links for nodes with wealth around 
$w$ in a given $dw$ interval) as a function of $w$, or by simply calculating 
$c(w,k)=\langle w\cdot k \rangle-\langle w \rangle  \langle k \rangle$. 
In Figures 9a and 9b we plot the values of $c(w,k)$ as
a function of time, and in Figure 9c we show the $ \langle k (w)\rangle$ curves. (For
constructing the curves in Figure 9c we used boxes of size $dw=0.1$.) 
From Figures 9a and 9b one notices 
again, that both the network structure and wealth distribution approach quickly (less than 5 MCS) a statistically stable limit. It is interesting to observe that the $c(w,k)$ correlation 
is stronger for low $p$ values, which makes sense since as $p$ 
increases the availability of
a wealthy node to accept more links decreases. 
As $p$ increases the $c(w,k)$ trend suggests that
we deal with a clear anti-correlation between the wealth and number of links of a node, which means that nodes which do not get too many links
will in general become wealthy. The trend of the 
$\langle k (w)\rangle$ curves (Figure 9c) suggests similar conclusions, but here we can also see this correlation
effect differentiated as a function of the $w$ value. In the low and medium wealth limit there is a
clear anti-correlation between wealth and number of links, while for the wealthy nodes (much fewer in number)
there is a positive correlation trend.  In Figure 9c. we plotted again the data only for
$w \le 5$, since for higher wealth values the curves are rather noisy due to poor 
statistics. 
   
\begin{figure}
\begin{center}
\epsfysize=70mm
\epsfxsize=130mm
\epsffile{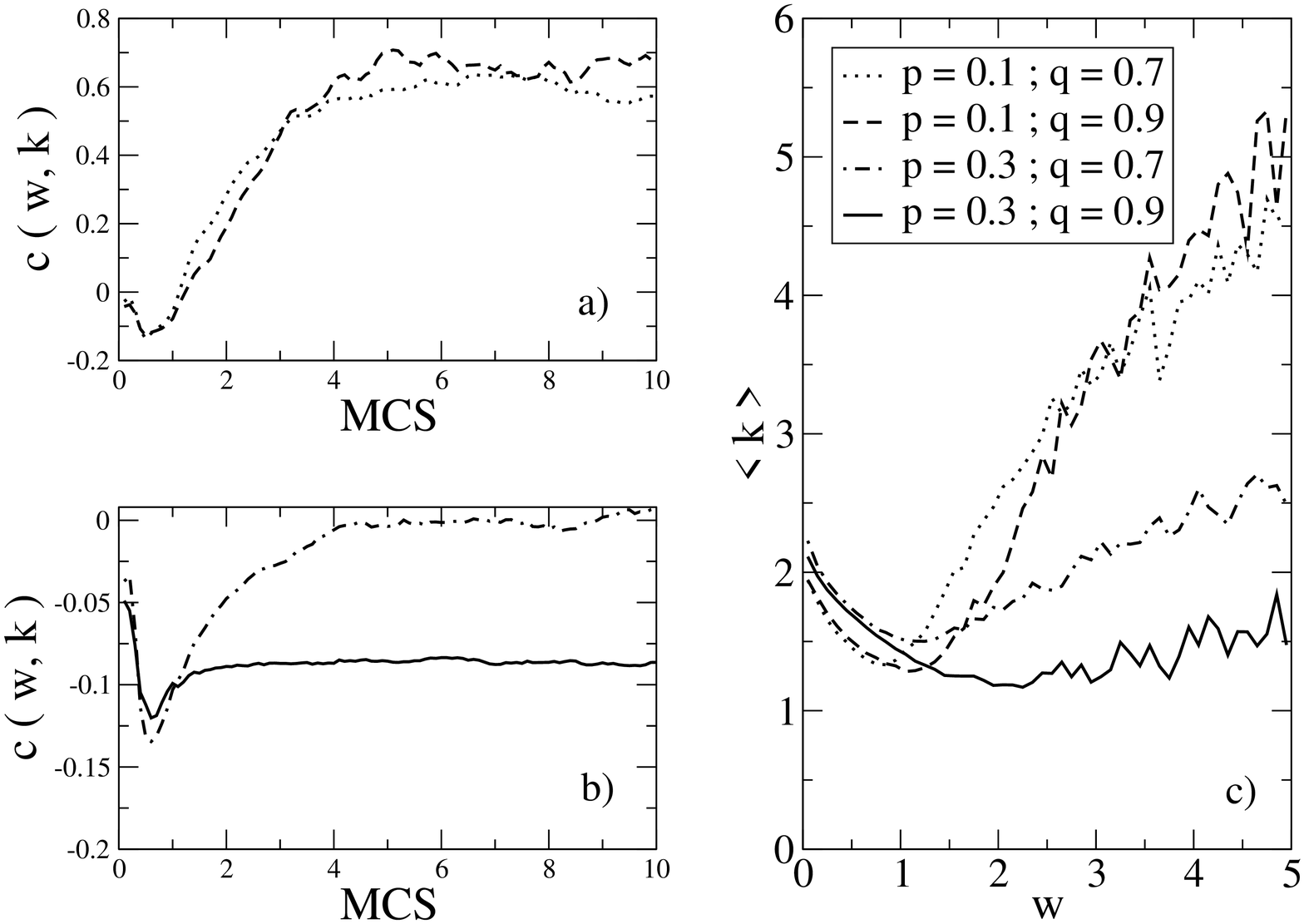}
\caption{Results for the correlation between the connectivity and wealth
of the nodes. Figure (a.) and (b.) shows results for $c(w,k)$ 
as a function of time considering the relevant $p$ and $q$ values. 
Figure (c.) illustrates the trend for the mean connectivity of nodes 
with different wealth values. For all the figures the corresponding
$p$ and $q$ values are given on the legend of (c), and we considered
$N=10000$ nodes.}
\label{fig9}
\end{center}
\end{figure}

\section{Discussion and comparison with real data}

Let us now analyze real wealth distribution data in societies in order to
check the quantitative agreement with our results.
We use estimates for the
distribution of personal wealth in United Kingdom 
(available on the Internet) \cite{UK}, based on inheritance tax, capital transfer tax and
other data (the methods used for getting these estimates are also described in \cite{UK}).
Plotting the cumulative wealth distribution for a chosen year (2001 in our case), 
one gets the graph in Figure 10 (distributions for other years are quite similar, even
quantitatively).

\begin{figure}
\begin{center}
\epsfysize=70mm
\epsffile{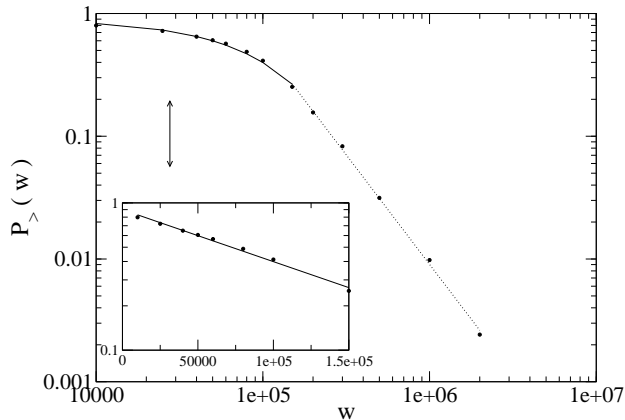}
\caption{Cumulative wealth distribution for the population of the
United Kingdom, for year 2001. Results obtained using the database from 
\cite{UK}. The power-law tail is described by an exponent $\alpha=1.78$. The inset
illustrates the initial exponential behavior of the curve, using a log-normal
scale}
\label{fig10}
\end{center}
\end{figure}

On the data presented in Figure 10, one can nicely identify the exponential regime for
low and medium wealth values, and the Pareto power-law distribution in the high
wealth limit. As emphasized in the introduction, the Pareto tail describes the upper
$5\%$ of the society. The UK-2001 data suggests a Pareto index $\alpha=1.78$ (Figure 10). An 
immediate comparison with the distribution obtained for our family-model (Figure 5), shows
that for the reasonable $q=0.7$ and $p=0.3$ parameters the model offers a fair description.
The Pareto index for these parameters is around $\alpha=1.8$, in the low and medium wealth limit the $P_>(w)$ curve is exponential, and the Pareto law is valid for the upper $5-10\%$ of the society. Concerning the shape of the $P_>(w)$ curve, the model thus seems to work well. 

The network structure generated by the model also seems to be realistic. The exponential
nature of the network, the most probable value of the connectivity $k_{prob} \approx 2$, and
the average connectivity $\langle k \rangle \approx 1.9$ are all reasonable for real first-degree family relation networks. 
The correlations $ \langle k_{nn} \rangle(k)$, $ \langle w_{nn} \rangle(w)$, 
$\langle w_{nt} \rangle(w)$ and
$\langle k \rangle (w)$, presented in Figures 7.-9., and described in the previous section, 
are also reasonable.  This kind of correlations could bee expected, since our model is
somehow similar to the ideas of hidden variables proposed by Bogu\~na et al \cite{bogunya04},
and their model also generated correlated networks.

Within the proposed model we can also identify the wealth-diversification mechanism that finally
leads to Pareto's law. The time evolution of the $c(w,k)$ correlations (Figure 9a and 9b), 
and the time-evolution for the $P_>(w)$ cumulative distribution functions (Figure 2) viewed 
in parallel give us important clues in this sense. In the beginning of the dynamics there is usually 
a strong anti-correlation effect ($d[c(w,k)]/dt<0$) between wealth and number of links. This means
that in this regime those nodes will become wealthier which have fewer number of links. The
Pareto tail here does not exist, and this is where the strong wealth-diversification starts. After this
initial transient regime, the $c(w,t)$ correlation will converge to a stable limit, and 
simultaneously the stable $P_>(w)$ cumulative distribution function with the 
Pareto tail is formed.
The main mechanism leading to the strong wealth-diversification in our model is thus the initial
strong anti-correlation between the wealth and the number of links of one node. 

One can also simply verify that the main necessary ingredient that will produce the power-law tail is
the preferential wealth redistribution in the system. Without the preferential wealth-redistribution
of the $2q$ amounts, the model will not generate power-law tails for $P_>(w)$. This rich-gets-richer
effect seems to be thus the main mechanism leading to power-law wealth distribution in the richer part of the societies.

\section{Conclusions} 

We have presented a family-network model designed to explain the cumulative wealth distribution
in societies. In our model the wealth-exchange is realized on a first-degree family relation network,
and it is governed by two parameters. The dynamics is defined through realistic 
rules and generates both the underlying family network and wealth distribution.
The model has a stable thermodynamic limit, and the dynamics quickly leads to a network structure and wealth-distribution which are stable in time. Extended computer simulations show, that for reasonable parameter 
values both the obtained cumulative wealth distribution function and network structure are realistic: 
(i) in good agreement with real measurement data we were able to generate cumulative 
wealth-distribution functions with Pareto-like power-law tails, (ii) the obtained Pareto index is close to the measured values, (iii) the cumulative wealth distribution function for the low and medium wealth values is exponential as found in social data (iv) the Pareto regime is valid for the upper $5\%$ of the society, (v) the generated first-degree family relation network is realistic.  We observed that in our model the initial wealth-diversification is realized through a strong anticorrelation between the wealth of the nodes and their number of links. As the main mechanism leading to the formation of the Pareto power-law tail we identified the preferential redistribution of wealth in the society. In the generated networks many interesting correlations have also been revealed. 

In spite of its strengths the proposed model is still a rough approximation to reality. One
may argue that many important cultural, social or economic phenomena have been neglected.
We consider this model as a first, mean-field type approximation. 
The novel aspect of our approach is however that the network structure was not a-priori
introduced in the model, but it got formed during the postulated wealth-exchange dynamics. 
Subscribing to the ideas presented in \cite{garlaschelli04b}, we also feel that such type of
approach should be considered for explaining many other social or economic phenomena and
complex network structures.

{\underline{Acknowledgment}}

Z. Neda acknowledges a Nato Fellowship and the excellent working atmosphere at the
Centro de F\'{\i}sica do Porto. R. Coelho was partially supported by a junior research grant (BI) from Centro de Física do Porto. J. Ramasco acknowledges a post-doctoral grant from FCT
(Portugal).

\end{document}